\documentstyle[12pt,epsfig]{article}

\begin{document}
\newcommand {\be}{\begin{equation}}
\newcommand {\ee}{\end{equation}}
\newcommand {\ba}{\begin{eqnarray}}
\newcommand {\ea}{\end{eqnarray}}
\newcommand {\bea}{\begin{array}}
\newcommand {\cl}{\centerline}

\newcommand {\eea}{\end{array}}
\renewcommand {\thefootnote}{\fnsymbol{footnote}}

\vskip .5cm

\renewcommand {\thefootnote}{\fnsymbol{footnote}}
\def \a'{\alpha'}
\baselineskip 0.65 cm
\begin{flushright}
SISSA/70/2002/EP \\
SLAC-PUB-9593\\
hep-ph/0211341 \\
\today
\end{flushright}
\begin{center}
{\Large{\bf On the Effective Mass of the Electron Neutrino in Beta
Decay \footnote{Work supported, in part, by U. S. Department of Energy,
under contract DE-AC03-76SF00515.}}}
{\vskip .5cm}
 Y. Farzan$^{1,2}$ and  A. Yu. Smirnov$^{3,4}$

$^1$ {\it Scuola Internazionale Superiore di
Studi Avanzati, SISSA, I-34014, Trieste, Italy}\\
$^2$ {\it Stanford Linear Accelerator Center, Stanford University,
2575 Sand Hill Road,
Menlo
Park, California 94025}\\
$^3$ {\it The Abdus Salam International Centre for
Theoretical Physics, I-34100 Trieste, Italy}\\
$^4$ {\it Institute for Nuclear Research, RAS, Moscow, Russia}\\

\end{center}
\begin{abstract}
In the presence of mixing between  massive neutrino states, the distortion
of the electron spectrum in  beta decay is, in general,
a function of several  masses and mixing angles.
For $3\nu$-schemes which describe the solar and atmospheric neutrino
data,  this distortion can be  described by a single
effective mass, under certain conditions.
In the literature, two different definitions  for the
effective mass have been suggested. We  show that for
quasi-degenerate mass schemes
 (with an overall mass scale $m$ and splitting $\Delta m^2$)
the  two definitions coincide up to $(\Delta m^2)^2/m^4$
corrections. We consider the  impact of different effective masses
on the integral energy spectrum.
We show that the spectrum with a single  mass can be used
also to fit the data in the case of  $4\nu$-schemes
motivated, in particular, by the LSND results. In this case
the accuracy of  the mass determination turns out to be better
than $(10 - 15)\%$.
\end{abstract}

\section{Introduction}
%%%%%%%%%%%%%%%%%%%%%%%%%%%%%%%%%%%%%%%%%%%%%%%%%%%%%%%%%%%%%%%%%%%

Determination of the absolute scale of neutrino  masses  is
one of the  most important and,  at the same time, challenging problems in
neutrino physics.
Currently, the study of  the electron energy spectrum
near the end point of the Tritium beta decay,
$^3_1{\rm H} \rightarrow ^3_2{\rm He}+ e^- +\bar \nu_e$,
is the most sensitive direct method of determining the
scale of masses.
It is well-known that in  the absence of mixing,
the energy spectrum of the emitted $e^-$ is described  by
\begin{equation}
{dN \over dE} = R(E)(E_0-E)[(E-E_0)^2-m_\nu^2]^{1 \over 2}\Theta
(E_0-E-m_\nu),
\label{mother}
\end{equation}
(see {\it e.g.} \cite{ketab}), where $E$ is the energy of
the electron, $E_0$ is the
total decay
energy and $R(E)$ is given by
\be
R(E)=G_F^2 {m_e^5  \over 2 \pi^3} \cos^2 \theta_C
|M|^2 F(Z,E)p E~.
\label{re}
\ee
Here $G_F$ is the Fermi constant, $p$ is the momentum of the electron,
$\theta_C$ is the Cabibbo angle and $M$ is the nuclear matrix
element.
$F(Z,E)$ is a smooth  function of energy which describes
the  interaction of the produced electron in the final state.
Both  $M$ and $F(Z,E)$ are independent of $m_\nu$, and therefore the
dependence of the
spectrum on $m_\nu$ follows only  from the phase space factors.

The analysis of the present data \cite{mainz,troitsk}
in terms of
Eq. (\ref{mother})  leads to the following bound:
$$
m_\nu<2.2 \ \ \ {\rm eV}.
$$
The forthcoming beta decay experiment, KATRIN~\cite{katrin},
with  energy resolution $\Delta E \sim 1$ eV   will be
sensitive to neutrino masses down to \be
m_{\nu} \sim 0.3 ~{\rm eV}.
\label{sens}
\ee

The  atmospheric and solar neutrino data provide  strong evidence
for neutrino mixing. In the presence of mixing, the electron
neutrino is a combination of
the mass eigenstates  $\nu_i$ with masses $m_i$: $\nu_e=\sum_i
U_{ei}\nu_i$.  In this case, instead of Eq. (\ref{mother}), the
expression for the spectrum is given by
\be
{dN_0 \over dE}=R(E) \sum_i |U_{ei}|^2
(E_0-E)[(E_0-E)^2-{m_i}^2]^{1 \over 2} \Theta (E_0-E-m_i),
\label{improve}
\ee
where the step function, $\Theta(E_0 - E-m_i)$,  reflects the fact that
a neutrino
can be produced only if the available  energy  is larger than its mass
\cite{shro}.
According to Eq. (\ref{improve}),
in general,  several  mass and mixing
parameters should be used to perform a fit of the  experimental data.

However, in the realistic situation a few
parameters (possibly even one) is
enough. In fact, the number and type of parameters
depend on:
(i) the type of the neutrino mass spectrum; (ii) the part  of
the spectrum being measured or
the width of the energy interval under consideration:
\be
\delta \equiv E - E_0~;
\ee
(iii) the  statistics  of the experiment;
(iv) the energy resolution of  the detector, $\Delta E$.
%and the width of the energy interval under consideration, $\delta$.
For the KATRIN experiment,  if there are only three
neutrinos and  at least one of their masses is in the
sensitivity range of KATRIN (see  Eq. (\ref{sens})), the
situation will be simpler. In this case
the spectrum
should be
strongly degenerate, {\it i.e.,} the mass splittings are too small to be
resolved ($\Delta m_{ij} \ll \Delta E \sim {\rm \ \ 1 \ \ eV} <\delta$).

The mass splitting is given by
$$
\Delta m_{ij} \equiv m_i - m_j = {m_i^2-m_j^2 \over m_i+m_j} \leq
\sqrt{\Delta m_{ij}^2}  \leq 0.05  \ \ {\rm eV},
$$
where the inequalities  hold for any type of  spectrum. The bound 0.05 eV
corresponds to the neutrino mass splitting
$\Delta m_{ij}^2 = \Delta m_{atm}^2 = 3 \times 10^{-3} $ eV$^2$.
For the range of sensitivity of KATRIN, $m_\nu \sim
0.3 - 2$ eV,
the mass splitting is even smaller:
$$
\Delta m  < 5 \times 10^{-3}  \ \ {\rm eV},
$$
and  correspondingly,
$$
\frac{\Delta m }{m} < 1.6 \times 10^{-2} .
$$
Since the energy resolution of the forthcoming
experiment, $\Delta E$, will be much larger than
$\sqrt{\Delta
m_{atm}^2}\sim 0.05$ eV, the experimental data will not be able to
resolve different ``kinks" associated with the different mass states
and,  in spite of the presence of mixing, the spectrum  can effectively be
described by Eq. (\ref{mother}) with a single effective mass $m_{\beta}$.
Hereafter we will refer to this approximation as the {\it single mass
approximation}.

In  Ref. \cite{vissani}, it has been shown that  for energies
$E_{\nu} = E_0 - E \gg m_{i}$,  the distortion of the  electron
energy spectrum  due to non-zero  neutrino mass and
mixing is determined by an  effective mass
\be
m_{\beta1}=\sqrt{\sum_i m_i^2 |U_{ei}|^2  }.
\label{mix1}
\ee
However, the  highest sensitivity
to  the mass of $\nu_i$ appears in the energy range  close to the end point
where $E_{\nu} \sim m_i$.
\footnote{In  Ref. \cite{vissani} it was also noted that  Eq.
(\ref{mix1}) gives a very  good fit  for $E_\nu \sim m_i$ too.}
% and the KATRIN experiment will concentrate on energies
%close to the end-point for which
%and the condition $E_\nu \gg m_i$ does not apply.
It was shown in   Ref. \cite{us}  that for $E_\nu \sim m_i$, the
integral spectrum can be described by the effective mass
\be
\label{ours}
m_{\beta 2}  = \sum_i m_i |U_{ei}|^2.
\ee
Recently, the  approximations based on the effective masses
$m_{\beta 1}$ and $m_{\beta 2}$
have been discussed in Ref. \cite{polish,strumia}.
It was claimed in Ref. \cite{polish} that for large energy
intervals, $\delta \gg m_i$, the
definition in Eq. (\ref{mix1})
provides a better fit of the  exact  beta decay spectrum,  and
consequently the mass  $m_{\beta 1}$ should be used in future analyses of the
experimental results.  We do not agree with this last statement.
In this paper we consider  the  issue  in more detail
and discuss various  aspects of the problem.
We also expand the analysis for $4\nu$-schemes.

The paper is organized as follows. In  Sect. 2,
we compare the two definitions of effective mass for the quasi-degenerate
mass
schemes. In  Sect. 3, we explore the effect of using each form of
effective mass on the integrated energy spectrum in quasi-degenerate
$3\nu$-schemes. In  Sect. 4, we perform the same analysis for
non-degenerate
$3\nu$-schemes.  In  Sect. 5, we address the question of whether it is
possible to fit
$4 \nu$-schemes by a single parameter, and  evaluate the corresponding
error.
Conclusions are presented in  Sect. 6.

 %%%%%%%%%%%%%%%%%%%%%%%%%%%%%%%%%%%%%%%%%%%%%%%%%%%%%%%%%%%%%%
\section{Effective masses for  quasi-degenerate neutrinos}
%%%%%%%%%%%%%%%%%%%%%%%%%%%%%%%%%%%%%%%%%%%%%%%%%%%%%%%%%%%%%

The  quasi-degenerate mass scheme
is the only $3\nu$-scheme that will cause an observable shift of
the end-point in   forthcoming $\beta$-decay experiments.
We show here that although the combinations in
Eqs. (\ref{mix1}) and (\ref{ours}) are
derived using completely different analyses,
they coincide up to corrections of order of $(\Delta m /m)^2$
for the quasi-degenerate mass scheme.

We represent  the neutrino mass eigenstates as
$m_i = m_j + \Delta m_{ij}$, where  $i = 1, 2, 3$ and $m_j$ can be any  of
the mass eigenstates. The expressions for $(m_{\beta1})^2$ and
$(m_{\beta2})^2$  can then be written as
\be
m_{\beta1}^2  = m_j^2 + 2 m_j \sum_i \Delta m_{ij}  |U_{ei}|^2 +
\sum_i (\Delta m_{ij})^2  |U_{ei}|^2
\label{mix1e}
\ee
and
\be
m_{\beta2}^2  = m_j^2 + 2 m_j \sum_i \Delta m_{ij}
|U_{ei}|^2 + \left(\sum_i \Delta m_{ij} |U_{ei}|^2\right)^2.
\label{mix2}
\ee
These two expressions  coincide up to the last terms, which are
of order   $(\Delta m_{ij})^2$. Furthermore, the difference turns
out to be even smaller  if one considers   realistic scenarios.
According to the experimental data, $\nu_e$ is mainly distributed
in two
states
(for definiteness, in $\nu_1$ and $\nu_2$) with mass splitting
$\Delta m_{12}^2 \equiv \Delta m_{sun}^2$ while  the contribution of the third
state (with splitting $\Delta m_{13}^2
= \Delta m_{atm}^2$) to $\nu_e$ is small:
$|U_{e3}|^2 <  0.04$. Taking $j = 1$ we obtain
$$
{m_{\beta1}^2 - m_{\beta2}^2 \over m_{\beta2}^2}
\sim
\frac{1}{4 m_1^4} {\rm Max}
\left[
\left(\Delta m_{sun}^2 \right)^2 |U_{e2}|^2,
\left(\Delta m_{atm}^2 \right)^2 |U_{e3}|^2 \right] <  10^{-5}.
$$
The numerical bound
corresponds to
$m_1 > 0.3$ eV. Clearly, this mass difference  is
unobservable.
This means that $m_{\beta1}$ and $m_{\beta2}$
are equally good for the description of the spectrum both
near the end point and far from it.

For comparison, we consider  another possible definition for the
effective mass:
\be
m_{\beta3} = m_1.
\ee
In this case according to (\ref{mix1e}),  the relative difference of mass
squared is  linear in  $\Delta m_{ij}$:
\be
{m_{\beta1}^{2} - {m}_{\beta3}^2 \over m_{\beta3}^2}
= 2 \sum_{i} |U_{ei}|^2 \frac{\Delta m_{i1}}{m_1}
\sim
\frac{1}{2 m_1^2} {\rm Max}
\left[\Delta m_{sun}^2  |U_{e2}|^2,
\Delta m_{atm}^2 |U_{e3}|^2 \right] < 10^{-3},
\ee
where the last number corresponds to  $|U_{e3}|^2 = 0.04$ and
$m_1 > 0.3$ eV. Even in this case the difference between the
effective masses is
negligible.

%%%%%%%%%%%%%%%%%%%%%%%%%%%%%%%%%%%%%%%%%%%%%%%%%%%%%%%%%%%%%%%%%%%%
\section{Effective masses and the integrated spectrum}
%%%%%%%%%%%%%%%%%%%%%%%%%%%%%%%%%%%%%%%%%%%%%%%%%%%%%%%%%%%%%%%%%%%%

The KATRIN experiment can operate in two modes \cite{intent}: i) as an
integrating spectrometer (MAC-E-FILTER mode);
ii) as a non-integrating spectrometer (MAC-E-TOF mode).
In the MAC-E-FILTER mode, the measured quantity is
the number of electrons in an interval $\delta$ close to the end-point.
For this mode, the energy resolution ($\Delta E \sim
1$ eV) is a measure of the sharpness of the filter.
In the MAC-E-TOF mode, the Time Of Flight of retarded electrons is
measured to derive the energy of the electron. The resolution
function is  triangular with a width of $\sim$ 1 eV.

KATRIN will operate primarily in the integrating mode and we therefore
focus here
on the
integrated  energy spectrum.
Let us study the error in the integrated energy spectrum caused by  using
the different definitions for the effective mass.
Since $R(E)$ in Eq. (\ref{re})
is a slowly varying function of energy
we can write the number of events above the energy
$  E_0 - \delta$
as
\be
\label{nofi}
n_a(\delta) =  \bar{R} \int_{E_0 - \delta}^{E_0}
{1 \over R}\frac{dN_a}{dE}dE, ~~~~(a = 0, 1, 2)
\ee
where the subscript $a$ identifies the distribution used;
$a = 0$ indicates the exact spectrum  in  Eq. (\ref{improve}), while
$a =
1,2$
indicate the
spectra with effective masses $m_{\beta 1}$ and $m_{\beta 2}$,
respectively. Here,
$E_0$ is the end-point of the spectrum for zero neutrino mass
and $\bar{R}$ is the averaged value of the function $R(E)$.

The integration in  Eq. (\ref{nofi})  leads to
\be
n_0(\delta) =  \frac{\bar{R}}{3}
\sum_i |U_{ei}|^2
\left(\delta^2 - m_i^2 \right)^{3/2},
\ \ \ \ \ \
n_a(\delta) =  \frac{\bar{R}}{3}
\left(\delta^2 - m_{\beta a}^2 \right)^{3/2}.
\label{nofo}
\ee

 Let us define the ratios
\be
\label{R}
r_a \equiv  {n_a - n_0   \over n_a},~~~ a = 1,2
\ee
 which  give the errors caused by the single mass approximation
(provided that the errors are  small).  Using Eq. (\ref{nofo}) we get
\be
\label{int}
r_a = 1 - {\sum_j |U_{e j}|^2 (\delta^2 - m_j^2)^{3/2}
\over (\delta^2 - m_{\beta a}^2)^{3/2}}.
\ee
We  introduce
\be
\Delta m_{ja} \equiv m_j - m_{\beta a};
\ee
as we have seen, for quasi-degenerate mass schemes
$\Delta m_{ja} \ll  m_{\beta a}$. Then, using the smallness of the ratio
\be
\frac{2 m_{\beta a} \Delta m_{ja}}{\delta^2 - m_{\beta a}^2  }
\ll 1,
\ee
we can expand expression Eq. (\ref{int}) as follows:
\be
\label{error}
r_a  =
\frac{3 m_{\beta a}}{\delta^2 - m_{\beta a}^2}
\sum_j |U_{ej}|^2 \Delta m_{ja} +
\frac{3 (\delta^2 - 2 m_{\beta a}^2)}
{2(\delta^2 - m_{\beta a}^2)^2}
\sum_j |U_{ej}|^2 (\Delta m_{ja})^2 +
{\cal O} \left(\frac{(\Delta m)^3}{\delta^3} \right).
\label{raa}
\ee

First we consider   the effective mass $m_{\beta 2}$,
{\it i.e.,}  $a = 2$. For $m_{\beta 2}$ the first term in
Eq. (\ref{error}) vanishes.
This can be  verified very easily;
$\sum_j |U_{ej}|^2 (m_{j} - m_{\beta 2}) =
\sum_j |U_{ej}|^2 m_j  - m_{\beta 2} = 0$,  where we have taken into
account the unitarity condition  $\sum_j |U_{ej}|^2 = 1$.
This cancellation   motivated us to
introduce the effective mass given by Eq. (\ref{ours}) in
Ref. \cite{us}.
As a consequence, for $m_{\beta 2}$ the deviation, $r_2$ is
of  ${\cal O}((\Delta m)^2)$:
\be
\label{error1}
r_2  = \frac{3}{2(\delta^2 - m_{\beta 2}^2)}
\left(1 - \frac{m_{\beta 2}^2}{\delta^2 - m_{\beta 2}^2} \right)
\sum_j |U_{ej}|^2 (\Delta m_{j2})^2~.
\ee
For $\delta \sim 1$ eV, it can be estimated as
\be
r_2 \sim \sum_j|U_{ej}|^2 \frac{(\Delta m_{j2})^2}{\delta^2}
\sim \frac{1}{(m\delta)^2}
{\rm Max}\left[(\Delta m_{sun}^2)^2 |U_{e2}|^2,
(\Delta  m_{atm}^2)^2|U_{e3}|^2 \right] < 4 \times 10^{-6}.
\label{estim}
\ee
Notice that the sum in   Eq. (\ref{error1}) can be written as
\be
\sum_j |U_{ej}|^2 (\Delta m_{j2})^2 =
\sum_j |U_{ej}|^2  m_{j}^2 - m_{\beta 2}^2 =
m_{\beta 1}^2 -  m_{\beta 2}^2.
\ee

For the effective mass $m_{\beta 1}$,  Eq.
(\ref{raa})  yields:
\be
\label{error2}
r_1  =  - \frac{3 m_{\beta 1}^2}{2(\delta^2 - m_{\beta 1}^2)^2}
\sum_j |U_{ej}|^2 (\Delta m_{j1})^2.
\ee
Again, the sum can be written in terms of effective masses, so that
\be
\label{error22}
r_1  =  - \frac{3 m_{\beta 1}^3 (m_{\beta 1} - m_{\beta 2})}
{(\delta^2 - m_{\beta 1}^2)^2}.
\ee
For $\delta = 1$  eV, $r_1$ and $r_2$  have the same  orders of magnitude
given
by Eq. (\ref{estim}).

According to Eqs. (\ref{error1}, \ref{error2}) the ratio of errors is
equal to
\be
\left|\frac{r_2}{r_1}\right| =
\frac{\delta^2 - 2m_{\beta}^2}{m_{\beta}^2}.
\ee
{}From this we see that
\be
r_2 < r_1 ~~~~~{\rm for} ~~~ \delta < \sqrt{3} m_{\beta}  \sim
\sqrt{3} m_1;
\ee
the approximation with $m_{\beta2}$  works better than the one
with $m_{\beta1}$ for $\delta$ not too large.
For  $\delta > \sqrt{3} m_{\beta}$ the spectrum with
$m_{\beta1}$ gives a better fit.

We now consider  the error  for large $\delta$.
Expanding  the general formula
in Eq. (\ref{raa}) in powers of $1/\delta^2$, we find that
 for $\delta \gg m_{\beta}$,
\be
\label{error4}
r_a  = \frac{3}{2\delta^2}
(\sum_j |U_{ej}|^2 m_{ja}^2 - m_{\beta a}^2 ) +
{\cal O}\left(\frac {m_\beta^4}{\delta^4}\right).
\ee
For  $m_{\beta 1}$, the corrections  of  order
$1/\delta^2$ vanish.  The explicit dependences of $r_a$
on $\delta$ are given by Eqs. (\ref{error1},\ref{error2}).
For $\delta \to \infty$,
$r_1 \propto  1/\delta^4$ and $r_2 \propto 1/\delta^2$;
as $\delta$ increases both approximations converge to
the exact result, however, the spectrum with $m_{\beta1}$ converges
faster than the one with $m_{\beta2}$.

We now use these considerations to  discuss the results
of ~\cite{polish}, where the quantity
\be
\label{theirs}
h(\delta, m_1)={|n_0(\delta, m_1)-n_2(\delta, m_1)| \over
|n_0(\delta, m_1)-n_1(\delta, m_1)|}
\ee
has been studied for different values of  $m_1$ (the lightest
neutrino mass).
For $m_1 = 0.1$ eV (a typical quasi-degenerate mass scheme) and small
values of $\delta $, the ratio $h(\delta, m_1)$ is smaller than one.
As $\delta$ increases, $h(\delta, m_1)$ diverges.
This  is the basis of statement in ~\cite{polish} that the spectrum with
$m_{\beta 1}$ gives a better approximation.

According to our results, for large $\delta$,
\be
h = \frac{r_2}{r_1} \propto \delta^2,
\ee
so that $h$ increases with   $\delta$.
However, this does not mean that the spectrum with
$m_{\beta2}$ fails to describe the exact integrated spectrum,
or that it is worse than the one corresponding to $m_{\beta1}$.
In fact, for the quasi-degenerate mass spectrum in Eq. (\ref{theirs}) both
the numerator and denominator are very small ( see
Eq. (\ref{estim})), and therefore the ratio
$h$ has no real meaning.

For large  $\delta$ the difference of the two approximations
is unobservable in spite of the possible accumulation
of absolute values of deviations. Let us consider this
in more detail.
%%
%In the vicinity of the end-point, the number of the detected electrons
%($\int (dN_0/dE)dE$) is suppressed so, the statistical uncertainty is
%large, and moreover near the end point
%$m_{\beta1}$-approximation is better.
%%
According to  Eq. (\ref{nofo}), the total number of events increases
with
$\delta$ since $n_a \propto \delta^3$. On the other hand,
the difference of numbers of events increases as
\be
n_2 - n_0 = n_2 \cdot r_2 \propto  \delta.
\ee
The statistical errors increase faster,
$\sqrt{n_a} \propto \delta^{3/2}$,
and therefore,
the sensitivity of the experiment to the deviation of the
$m_{\beta2}$-approximation from exact spectrum (\ref{improve}) decreases with
the increase of $\delta$. Similarly we find
$n_1 - n_0  \propto  1/\delta$, so that even the deviation in
absolute number of events  decreases for the effective mass $m_{\beta 1}$.
Furthermore,  if the two approximations are indistinguishable near the
end point it is not possible to distinguish them
by increasing the integration region. Even in the case that the absolute
difference of events increases, the statistical error increases  faster.

%%%%%%%%%%%%%%%%%%%%%%%%%%%%%%%%%%%%%%%%%%%%%%%%%%%%%%%%%%%%%%%%%%%%%%%
\section{Non-degenerate $3\nu$-schemes}
%%%%%%%%%%%%%%%%%%%%%%%%%%%%%%%%%%%%%%%%%%%%%%%%%%%%%%%%%%%%%%%%%%%%%%%%

In this section
we consider schemes for which the heaviest mass
is of order  $
\sqrt{\Delta m^2_{atm}}
\sim 0.05 - 0.07$ eV, which is far below the reach  of KATRIN experiment.
Therefore, the discussion here  is relevant only for  hypothetical
future experiments with substantially better sensitivity and
higher  energy resolution.
We expect that for  $\beta$-decay
experiments in the near future, $\Delta E, \delta \gg m_1,m_2$.
Consequently, using Eq. (\ref{int}) we
can write
\be \label{new}
r_a=\frac{3}{2}{\sum_i m_i^2|U_{ei}|^2-m_{\beta a}^2 \over
\delta^2}-\frac{3}{8}{\sum_i m_i^4 |U_{ei}|^2 \over
\delta^4}-\frac{15}{8}{m_{\beta a}^4 \over \delta^4}+\frac {9}{4}{m_{\beta
a}^2 \Sigma_i m_i^2|U_{ei}|^2 \over \delta^4}.
\ee

We first  study  schemes for which $m_1\simeq m_2\sim \sqrt{\Delta
m_{atm}^2}$ (we remind the reader  that the electron-neutrino is mainly
distributed in
the 1- and 2-states with  $|m_1^2-m_2^2|=\Delta m_{sun}^2.$) Note that this
includes the inverted hierarchical scheme as well as the normal scheme
with a
non-zero lightest mass.
If in  Eq. (\ref{new}) we set  $m_{\beta a}$ equal to either $m_{\beta
2}$,
 $m_1$ or $m_2$,
the error will be of order
$$
{\rm Max}\left[ \left( {\Delta m_{atm}^2  \over
\delta^2} \right)|U_{e3}|^2,\left({\Delta m_{atm}^2 \over\delta^2}
\right)^2,\left({\Delta m_{sun}^2 \over\delta^2}\right)\right],
$$
which will probably be smaller than the experimental errors;
it is  safe
to use $m_{\beta 2}$. However, $m_{\beta 1}$ gives an even  better fit:
$$
r_1\sim {\rm Max}\left[ {(\Delta m_{sun}^2)^2 \over \delta^4},({\Delta
m_{atm}^2 \over \delta^2}) ^2|U_{e3}|^2 \right],
$$
where we have assumed that $m_3$ is also of the order of
$\sqrt{m_{atm}^2}$.

We now  discuss the normal  hierarchical mass scheme, $m_1\ll m_2\simeq
\sqrt{\Delta m_{sun}^2} \ll
m_3\simeq \sqrt{\Delta m_{atm}^2}$, neglecting the effect of the lightest
state.
If $|U_{e3}|$ is close to its present
upper bound, we can also  neglect  the second state,
so that
\be
m_{\beta1} \approx m_3 |U_{e3}|,~~~~m_{\beta2} \approx m_3 |U_{e3}|^2.
\ee
In this case $(m_{\beta1} - m_{\beta2})/m_{\beta1} \sim 1$, and in
fact,
$m_{\beta1} \gg m_{\beta2}$.  The
effect of the neutrino mass would consist of a kink
at  $ E=E_0-m_3$
whose size is determined by $|U_{e3}|^2$.
For the  realistic case  $\delta  \gg  m_3$, the statistics should
be very high to detect the deficit of the total number of events
above $E_0 - \delta$.
In this case we can  use a single effective mass instead of
several parameters; the error $r_a$ defined in  Eq. (\ref{R}),
is negligible
for an appropriate choice of effective mass.
For $m_{\beta 1}$, the first term in  Eq. (\ref{new}) vanishes, and the
$m_{\beta 1}$ approximation gives a very good description of the data.
In contrast,
$m_{\beta 2}$ does not describe the situation well; the deviation of
the $m_{\beta 2}$ approximation  from the exact spectrum is comparable
to the effect of non-zero mass itself.

%%%%%%%%%%%%%%%%%%%%%%%%%%%%%%%%%%%%%%%%%%%%%%%%%%%%%%%%%%%%%%%%%%%%%%%%%%%
\section{Four-neutrino mass schemes}
%%%%%%%%%%%%%%%%%%%%%%%%%%%%%%%%%%%%%%%%%%%%%%%%%%%%%%%%%%%%%%%%%%%%%%%%%%%%%
In this section we study the  effective mass approximation
for  the case of 4$\nu$-schemes  motivated by the
results of the LSND experiment
\cite{no4}.
(Apart from the LSND results, there are other motivations for the
existence of  a 4th neutrino.
One can use the arguments described below
for more general cases of active-sterile neutrino mixing.)

The main feature of these schemes is  the existence of {\it two groups} of
states
separated by a large mass gap given by  $\Delta m_{LSND}^2\sim 1 $
eV$^2$.
The mass splitting within each group is small: $\Delta m^2  \ll  \Delta
m_{LSND}^2$.
\footnote{Clearly, there is  no splitting if the group  consists only of
one
state.}
Furthermore, the admixture of $\nu_e$  in one of these groups is very
small;
it is restricted by the  reactor experiments CHOOZ and BUGEY.

In  3+1 schemes $\nu_e$ is mainly distributed in three mass
eigenstates with small mass differences given by $\Delta m_{sun}^2$ and
$\Delta m_{atm}^2$, while the fourth eigenstate  (separated by a
mass gap $\Delta m_{LSND}^2$ from the
rest of the states) has a small admixture of the  electron neutrino, {\it
i.e.,}
$|U_{e4}|\ll 1$.

In  2+2 scheme, $\nu_e$ is mainly distributed in the first and second
mass eigenstates with a small splitting given by $\Delta m_{sun}^2$. The
third and fourth mass eigenstates, separated from the rest of the
states by $\Delta m_{LSND}^2$, have
small admixtures of $\nu_e$  given  by $|U_{e3}|^2$ and $|U_{e4}|^2$.

%$|U_{e3}|^2+|U_{e4}|^2$.
%Again, if the third and fourth mass eigenstates are heavier, we call the
%scheme normal.  Otherwise, it is called inverted scheme.

We identify the  groups of mass eigenstates  as $h$- (heavy)
and $l$- (light) groups.
As  discussed in  Ref. \cite{us}, the effect of each group
on the beta decay spectrum
%which are separated by $\Delta m_{LSND}^2$
can be described by only two parameters:
\be
\rho_{a} = \Sigma_i |U_{ei}|^2 ~~~{\rm and}~~~
m_a =\Sigma_i m_i|U_{ei}|^2/\rho_a,   ~~~(a = h, l),
\ee
where $i$ runs over the members of  each group and $m_h^2 - m_l^2 \simeq
\Delta m_{LSND}^2$.
Unitarity implies $\rho_l + \rho_h =1.$
The effect of the heavier group is a kink whose size and position are
given by $\rho_h$ and $m_h$, respectively.
The lighter group leads to a shift of  the end-point.
As for the case of the $3\nu$-scheme it can be shown that the error due to
using the
effective mass parameters $m_a$ is negligible.

We denote
\be
\rho_e \equiv {\rm Min} [\rho_h, \rho_l];
\ee
 in the (3 + 1) scheme $\rho_e = |U_{e4}|^2$,  while in the
(2 + 2) scheme  $\rho_e = |U_{e3}|^2+|U_{e4}|^2$.
In the range of $\Delta m_{LSND}^2 \sim (0.3 - 2)$ eV$^2$,
the strongest upper bound on  $\rho_e$ follows from the BUGEY experiment
\cite{bugey}:
\be
\rho_e < 0.027~~~(90 \%) ~~~ {\rm C.L.}
\label{b-rho}
\ee
In what follows we will refer to the scheme in which
 the heavy set has a small admixture of $\nu_e$, $\rho_h =  \rho_e$,  as
``normal" scheme.
The scheme in which
the light group has the smaller contribution to $\nu_e$,  $\rho_l =
\rho_e$, will be called the ``inverted" scheme.

For  the normal schemes   (either 2+2 or 3+1),
the beta  spectrum should have a ``small" kink
at $E_0- m_h$, with a height characterized by $\rho_e$
and the end point will be shifted to  $E_0- m_l$. In contrast,
for the inverted schemes  we expect
a ``large" kink of size  $1 - \rho_e \approx 1$ at $E_0- m_h$,
and  a small ``tail" after the kink
with a height $\propto \rho_e$.
In general, the spectrum with a given type of hierarchy is described by
three parameters:
$\rho_e$, $m_h$, and $m_l$.

Suppose {\it a priori} we do not know about the existence of the 4th
neutrino, and therefore
try to fit the integrated beta  spectrum using the single mass
approximation:
\be
n_s(\delta) = \frac{\bar R}{3} (\delta^2 - m_f^2)^{3/2},
\label{single}
\ee
where $m_f$ is the  fit mass parameter.
We clarify below the meaning of such a fit.

1) We consider first the  {\it normal} mass
scheme.
Due to the smallness of $\rho_e$, the main effect will  be produced by
$m_l$.
The single mass fit (Eq. (\ref{single})) means that we neglect
the kink and set  $m_f \simeq m_l$.
Let us estimate the error due to neglecting the kink,   $(m_f - m_l)/m_l$.

If $\delta < m_h$, the energy interval  does not contain
the kink, and we can write
\be
n(\delta)  = \frac{\bar R}{3}(1 - \rho_e)(\delta^2 - m_l^2)^{3/2}.
\label{truesmall}
\ee
Then, fitting  $n(\delta)$ with $ n_s(\delta)$  (that is, equating $n(\delta) =  n_s(\delta)$)
we find
\be
{m_f - m_l \over m_l} \approx \frac{\rho_e}{3}{(\delta^2 - m_l^2) \over
m_l^2 },
\label{mx2}
\ee
where we have used the smallness of $\rho_e$.
According to this expression, the mass $m_f$ which we
obtain from the single mass fit is larger than the
exact mass  $m_l$. The relative mass difference increases
with $\delta$
and decreases with
$m_l$. For $m_l = 0.3$ eV and $\delta = 1$ eV, using the bound
in  Eq. (\ref{b-rho}),  we obtain
$(m_f - m_l)/ m_l <  0.1$.  That is the error is smaller than $10\%$.

If $\delta > m_h$ ({\it i.e.,} the energy interval contains
the kink),  the number of events amounts to
\be
\label{kink}
n(\delta)=\frac{\bar R}{3}\left[\rho_e(\delta^2 - m_h^2)^{3/2}
+(1-\rho_e)(\delta^2 - m_l^2)^{3/2}\right].
\ee
Fitting $n(\delta)$ with the approximate spectrum $n_s(\delta)$
in  Eq. (\ref{single}), we obtain
\be
\label{mx1}
{m_f - m_l \over m_l} = \frac{\rho_e}{3}
{(\delta^2-m_l^2)^{3/2} - (\delta^2-m_h^2 )^{3/2}
\over (\delta^2-m_l^2)^{1/2}m_l^2},
\ee
where we have neglected the higher orders of $\rho_e$.
The relative difference increases with $\delta$ and
in the limit $\delta \gg m_h$ asymptotically approaches  to
\be
\label{mx1lim}
{m_f - m_l \over m_l} = \frac{\rho_e}{2} {\Delta m^2_{LSND} \over m_l^2}.
\ee
For $\Delta m^2_{LSND} = 1$ eV$^2$ and  $m_l = 0.3$ eV using  Eq.
(\ref{b-rho}), we find that $(m_f - m_l)/ m_l < 15~~ \%$.\\
This can also be interpreted  in this way: the effect of the small
kink can be resolved only if the experimental errors are smaller than
$\sim 15$ \%.

2) We now  consider  the  inverted schemes. In these cases the main
effect
is a large kink at $E_0-m_h$. Fitting the spectrum with a single
mass,
we are not sensitive to the small tail after the kink and  the fit
parameter $m_f$
corresponds to
$m_h$. Let us evaluate the relative difference of $m_f$ and   $m_h$.
The exact spectrum is
\be
\label{kink2}
n(\delta) = \frac{\bar R}{3} \left[\rho_e(\delta^2 - m_l^2)^{3/2}
+(1-\rho_e)(\delta^2 - m_h^2)^{3/2}\right].
\ee
(It corresponds to $n(\delta)$ for the normal scheme substituting
$h \leftrightarrow l$.) Fitting $n(\delta)$ with  $n_s(\delta)$, we
obtain:
\be
\label{mx3}
{m_h - m_f \over m_h} = \frac{\rho_e}{3}
{(\delta^2-m_l^2)^{3/2} - (\delta^2-m_h^2 )^{3/2} \over (\delta^2-m_h^2)^{1/2}m_h^2}.
\ee
Notice that for the inverted schemes $m_h > m_f$. The absolute value
of the relative
difference, as  in the case of the normal scheme, is limited by the
quantity in the right-hand side of Eq.
(\ref{mx1lim}).

Most probably, KATRIN will start  operation, after   results of the
MiniBooNE experiment~\cite{boone} are released.
If MiniBooNE does not see any oscillation effect, a stronger bound on
$\rho_e$ will be obtained
and any possible effect of a fourth neutrino on the beta spectrum will
be further
suppressed.
If  the MiniBoone experiment confirms the  LSND results,
the existence of a 4th neutrino will be confirmed and
the values of $\Delta m_{LSND}^2$ and $\rho_e$ will be determined.
%%So, KATRIN needs to fit only one mass parameter $m_{l}$ or $m_h$.
However,
MiniBooNe  will not be able
to discriminate between the normal and inverted schemes so, {\it a priori}
we
do not
know which formula, Eq. (\ref{kink}) or Eq. (\ref{kink2}),
describes the exact spectrum.

KATRIN not only can determine the neutrino mass scale, but also can help
to
discriminate between normal and inverted schemes.
In this direction, the following strategy can be used:

First, a fit of the data using  single mass approximation
given by  (\ref{single}) can be performed.
Then the value of the fit  parameter
$m_f^2$ should  be compared to  $\Delta m_{LSND}^2$.
If $m_f^2  < \Delta m_{LSND}^2$, considering the fact that
$m_h \geq \sqrt{\Delta m_{LSND}^2}$ we can
conclude that the scheme is  normal
and that $m_f \approx m_l$. To extract the value of $m_l$
 more accurately  (if the experimental uncertainties allow), we
can invoke  Eq. (\ref{kink}).
If $m_f^2 \geq  \Delta m_{LSND}^2$, the situation is ambiguous. In this
case, either of the following schemes is possible:
\newline
a) an inverted  mass scheme with $m_h^2 = m_f^2$ and
$m_l^2 = m_f^2 - \Delta m_{LSND}^2$, or
\newline
b) a normal mass hierarchy with $m_l^2 = m_f^2$ and $m_h^2 = m_f^2 +
\Delta m_{LSND}^2$.
\newline
In principle, this ambiguity can be solved by a non-integrating
spectrometer.
If a small kink around $E_0-\sqrt{m_f^2+\Delta m_{LSND}^2}$ is observed,
the
scheme is normal.
\newline
Note that for $m_f^2 > \Delta m_{LSND}^2$, the error ($(m_l-m_f)/m_f$ for
the normal schemes and $(m_h-m_f)/m_f$
for the inverted schemes) is less than 0.1 \%, which is negligible.
If the accuracy of a hypothetical integrating spectrometer is better than
this, by studying the dependence of $m_f$ on $\delta$ we can determine
whether the scheme is  normal
or inverted.
%%%%%%%%%%%%%%%%%%%%%%%%%%%%%%%%

Recently to reconcile the LSND results with
the results of the other neutrino
experiments another scenario has been
suggested \cite{cpt}.
In this scenario, there is no need for
sterile neutrinos but instead,  the masses
of
neutrinos and anti-neutrinos are different.
It is supposed that this phenomenological
model originates from  a more fundamental
non-local theory in which CPT is violated
in the neutrino sector \cite{theo}.
Note that the Tritium beta-decay experiments can
only
provide information about the masses of
anti-neutrinos because in these experiments
only the electron anti-neutrino is
involved.

In Ref. \cite{kamkam}  a model is
proposed which can  accommodate the LSND results as well as the results
of KamLAND
and other neutrino experiments.
In this model the anti-neutrino sector is composed of
three mass eigenstates with splittings
$\Delta m_{Kam}^2 \sim 10^{-4} \ \ {\rm
eV}^2$ and $\Delta m_{LSND}^2 \sim 1 \ \
{\rm eV}^2$. The electron anti-neutrino is
mainly distributed  in the states with the
small splitting ($\Delta m_{Kam}^2$) while
the contribution of the third state to
$\bar \nu_e$ is bounded by Bugey
($|U_{e3}|^2<0.027$).
Since KATRIN will not be able to
resolve the small splittings $\Delta
m_{Kam}^2$, $\Delta m_{LMA}^2$ and $\Delta
m_{atm}^2$, the signature of the
CPT-violating model described in Ref \cite{kamkam} will be exactly the
same as
the four-neutrino schemes and any result we
have found above applies
to this model as well.

 %%%%%%%%%%%%%%%%%%%%%%%%%%%%%%%%%%%%%%%%%%%%%%%%%%%%%%%%%%%%%%%%%%%%%%
\section{Conclusions}
%%%%%%%%%%%%%%%%%%%%%%%%%%%%%%%%%%%%%%%%%%%%%%%%%%%%%%%%%%%%%%%%%%%%%%
We have discussed the application of the effective mass approximation to
the energy spectrum of beta decay for different mass schemes, and  have found the following
results.

1) For a quasi-degenerate $3\nu$-scheme, which is the only scheme that can
have an observable effect on KATRIN, both definitions of the
effective mass, $m_{\beta 1}$ (\ref{mix1}) and  $m_{\beta 2}$
(\ref{ours}),
give a very good fit and are practically indistinguishable.  In the   analysis of the KATRIN data
one can safely use a single effective mass,
which can be identified  with any of $m_{\beta 1}$, $m_{\beta 2}$ or
$m_{\beta 3}$. The accuracy of the forthcoming measurements
will not be enough to distinguish between the different mass definitions.
We have found that
increasing the integration region  decreases the sensitivity to
the mass.  Notice that if in the future, measurements
 close to the end point ($\delta < 2 m_1$) with very high sensitivity
become  possible, $m_{\beta 2}$  will provide a better fit.

2) We have discussed the effect of non-degenerate $3\nu$-schemes on
measurements of hypothetical new experiments  sensitive to
neutrino mass down to $\sim \sqrt{\Delta m_{atm}^2}$. For an inverted
hierarchical $3\nu$-scheme, the
discussions and conclusions are similar to the quasi-degenerate case.
In the case of a normal mass hierarchy,
$m_{\beta 1}$ provides a good fit for the integrated energy spectrum
with $\delta=E-E_0\gg m_{\beta 1}$. However, the $m_{\beta 2}$
approximation
fails in this case, in the sense that the deviation is comparable to
the mass effect itself.

3) For the 4$\nu$-schemes motivated by the LSND results,
 non-zero masses will lead to  both a kink in the spectrum shape and a
shift of the end-point in beta
spectrum \cite{us}.
We show that, due to smallness of $\rho_e$
(the admixture of the electron neutrino in one of the groups of  mass
eigenstate)
 the data can be fitted by a spectrum with a single mass $m_f$ to a good
approximation.
For $m_f > 0.3$ eV (that is, in the range  of the KATRIN sensitivity)
this mass  equals to the  exact mass  to
better than $(10 - 15) \%$ accuracy.

In summary, in this letter we have
clarified the question that
which kind of  theoretical spectrum should
be used to  fit the experimental data on beta decay.
We have shown that the present knowledge on the
oscillation parameters
gives rather definite answer to this question which depends on
the sensitivity
of the experiment to neutrino mass and on the
number of neutrino
mass eigenstates involved.
For $3\nu$-schemes we have  shown that to
analyze the results of
experiments with sensitivity around that of  KATRIN,
one can  fit the spectrum with a single
mass $m_{\beta}$ identified by any of the  combinations
$m_{\beta i}$,  $i = 1,2,3$.
In the case of four (or more) neutrino
eigenstates, it is also possible to use the single mass approximation.
In this case $m_{\beta}$
gives the effective mass of the heavy or light group (depending on the
type of
the mass scheme)
of the quasi-degenerate states with an accuracy better than
20\%.

\section*{Acknowledgments}
We would like to thank F. Vissani for helpful comments. We are also
grateful to F. Petriello and H. Quinn for fruitful comments and
careful
reading of the
manuscript.

\end{document}